\documentclass[twocolumn,showpacs,preprintnumbers,aps]{revtex4}
\usepackage{amsmath}
\usepackage{mathrsfs}
\usepackage{dcolumn}
\usepackage{bm}
\usepackage{graphicx}
\usepackage{amsfonts}
\usepackage{amssymb}
\usepackage{bbm}
\usepackage{float}
\usepackage[dvipdfm]{hyperref}

\begin{document}

\title{Quantum speedup in a memory environment}
\author{Zhen-Yu Xu $^{1}$}
\email{zhenyuxu@suda.edu.cn}
\author{Shunlong Luo $^{2}$}
\email{luosl@amt.ac.cn}
\author{W. L. Yang $^{3}$}
\author{Chen Liu $^{1}$}
\author{Shiqun Zhu $^{1}$}
\email{szhu@suda.edu.cn}
\affiliation{$^{1}$School of Physical Science and Technology, Soochow University, Suzhou
215006, China}
\affiliation{$^{2}$Academy of Mathematics and Systems Science, Chinese Academy of
Sciences, Beijing 100190, China}
\affiliation{$^{3}$State Key Laboratory of Magnetic Resonance and Atomic and Molecular
Physics, Wuhan Institute of Physics and Mathematics, Chinese Academy of
Sciences, Wuhan 430071, China}

\begin{abstract}
Memory (non-Markovian) effect is found to be able to accelerate quantum
evolution [S. Deffner and E. Lutz, Phys. Rev. Lett. \textbf{111}, 010402
(2013)]. In this work, for an atom in a structured reservoir, we show that
the mechanism for the speedup is not only related to non-Markovianity but
also to the population of excited states under a given driving time. In
other words, it is the competition between non-Markovianity and population
of excited states that ultimately determines the acceleration of quantum
evolution in memory environments. A potential experimental realization for
verifying the above phenomena is discussed by using a nitrogen-vacancy
center embedded in a planar photonic crystal cavity under current
technologies.
\end{abstract}

\pacs{03.65.Yz, 37.30.+i, 03.67.-a, 42.50.-p}
\maketitle


\section{Introduction}

When it comes to the ultimate physical limit of a computer, two questions
arise immediately: how much information can it store, and how fast can it
run \cite{ultimate limits}? The former is related to the famous Boltzmann's
entropy formula that restricts the number of (qu)bits available to a
physical system \cite{Book QIC}. The latter concerns the speed of quantum
evolution, and is limited by the remarkable time-energy uncertainty
relation: $\tau \geq \max \left\{ \pi \hbar /(2\Delta E),\pi \hbar
/(2E)\right\} ,$ with the two bounds usually referred to as Mandelstam-Tamm
\cite{MT} and Margolus-Levitin \cite{ML} types, respectively. This
fundamental relation determines that it takes at least $\tau _{\mathrm{QSL}%
}=\max \left\{ \pi \hbar /(2\Delta E),\pi \hbar /(2E)\right\} $, the
so-called quantum speed limit (QSL) time, to evolve to an orthogonal state
for a quantum system with energy spread $\Delta E$ or average energy $E$.
This QSL time has been proved to be tight \cite{unified}, can be extended to
nonorthogonal state cases \cite{fz}, and plays an important role in a great
deal of quantum tasks such as quantum computation \cite{computation},
quantum communication \cite{state transfer}, quantum optimal control \cite%
{optimal control}, and quantum metrology \cite{metrology}.

Due to the inevitable coupling to the surrounding environment, quantum
systems are usually treated as open \cite{Book-Open}. Significant
experimental progress has been made in controlling the dynamics of open
systems \cite{opEion,opEphoton,opEsolid}. How to detect the limit to quantum
non-unitary evolution of open systems now becomes paramountly important. In
recent years, efforts have been made towards the description of QSL time in
the evolution of open systems \cite{op1,op3}. A unified lower bound,
including both Mandelstam-Tamm and Margolus-Levitin types, has been derived
by Deffner and Lutz \cite{op3}. By von Neumann trace inequality and
Cauchy-Schwarz inequality, the QSL time between an initial open system state
$\rho =\left\vert \psi _{0}\right\rangle \left\langle \psi _{0}\right\vert $
and its target state $\rho _{\tau }$, governed by the time-dependent
non-unitary equation $\dot{\rho}_{t}=\mathcal{L}_{t}\rho _{t}$ with $%
\mathcal{L}_{t}$ a super-operator, is given by
\begin{equation}
\tau _{\mathrm{QSL}}=\max \left\{ \tau _{1},\tau _{2},\tau _{\infty
}\right\} ,  \label{QSL time}
\end{equation}%
where $\tau _{p}=\sin ^{2}\left[ B(\rho ,\rho _{\tau })\right] /E_{\tau
}^{p}, $ $E_{\tau }^{p}=\frac 1{\tau} \int_{0}^{\tau }dt\left\Vert \mathcal{L%
}_{t}\rho _{t}\right\Vert _{p},$ and $\left\Vert A\right\Vert _{p}=(\alpha
_{1}^{p}+\cdots +\alpha _{n}^{p})^{1/p}$ denotes the Schatten $p$-norm, $%
\alpha _{1},\cdots ,\alpha _{n}$ are the singular values of $A$, and $B(\rho
,\rho _{\tau })=\arccos \sqrt{\left\langle \psi _{0}\right\vert \rho _{\tau
}\left\vert \psi _{0}\right\rangle }$ denotes the Bures angle between the
initial pure state $\rho =\left\vert \psi _{0}\right\rangle \left\langle
\psi _{0}\right\vert $ and the target state $\rho _{\tau }$.

To study the environmental effects on QSL time, one method is to evaluate
the characteristic of the intrinsic speed of the quantum evolution, i.e.,
given a driving time, how fast can a quantum system evolve. Interestingly,
it is discovered that non-Markovian effect can speed up the quantum
evolution with a damped Jaynes-Cummings model for an atom resonantly coupled
to a leaky single mode cavity: QSL decreases when the non-Markovian effect
becomes stronger \cite{op3}. Several questions, however, naturally arise:
(i) is the speedup phenomenon a rather general feature which also exists in
other physical models? (ii) is non-Markovianity the only key factor for
speeding up quantum evolution? (iii) what is the mechanism for quantum
acceleration in memory environments?

To address these questions, in this paper, we consider a two-level atom
embedded in a photonic crystal cavity (PCC) with the periodic dielectric
structures forming photonic band gaps (PBG) \cite{Book-PBG,PBG-review}. This
structured reservoir has been widely used to form atom-photon bound states
\cite{PBG-inhibition,PBG}, strong localization of light \cite{PBG-s}, as
well as entanglement preservation \cite{PBG-esd}. In this setting, the
transition from no-speedup to speedup of quantum evolution is observed when
the atomic frequency is approaching and going inside the band edge. QSL time
is found to be related to two quantities: non-Markovianity during the
driving time and excited population at $\tau $. We illustrate that it is the
competition between the two quantities that ultimately determines the
speedup of quantum evolution. Finally, a possible experimental realization
for our illustrated phenomena by a single nitrogen-vacancy (N-V) center
embedded in a two-dimensional PCC is discussed.

The work is organized as follows. In Sec. II, we introduce the physical
model with a two-level atom embedded in a PCC and derive the relevant QSL
time in Sec. III. In Sec. IV, we explore the mechanism for the speedup of
quantum evolution and suggest a PCC-N-V based experimental proposal in Sec.
V. Finally, we discuss and summarize the results in Sec. VI.

\section{Physical model}

A two-level atom (open system) embedded in a PCC is coupled to the radiation
field (environment) with the Hamiltonian ($\hbar =1$) \cite{Book-Open}
\begin{equation}
H=\omega _{0}{\sigma }_{+}{\sigma }_{-}+\sum_{k}\omega _{k}{a}_{k}^{\dag }{a}%
_{k}+i\sum_{k}g_{k}\left( {a}_{k}^{\dag }{\sigma }_{-}-{a}_{k}{\sigma }%
_{+}\right) ,  \label{H}
\end{equation}%
where $\omega _{0}$ is the resonant transition frequency of the atom between
the excited and the ground states, ${\sigma }_{\pm }$ are the Pauli raising
and lowering operators, $\omega _{k}$ and ${a}_{k}({a}_{k}^{\dag })$ are,
respectively, the frequency, the annihilation and creation operators of the $%
k$th mode of the reservoir with the real coupling constant $g_{k}=\omega
_{0}(2\epsilon _{0}\omega _{k}V)^{-1/2}\mathbf{e}_{k}\cdot \mathbf{d}$. Here
$\epsilon _{0}$ is the free space permittivity, $V$ and $\mathbf{e}_{k}$
refer to the normalized volume and the unit polarization vector of the
radiation field, respectively, and $\mathbf{d}$ is the dipole moment of the
atom.

The master equation for the reduced density matrix of the atom is $\dot{\rho}%
_{t}=\mathcal{L}_{t}\rho _{t}$ with
\begin{equation}
\mathcal{L}_{t}\rho _{t}=i\epsilon _{t}\left[ {\sigma}_{+} {\sigma}_{-},\rho
_{t}\right] +\gamma _{t}\left( {\sigma}_{+} {\sigma}_{-}\rho _{t}+\rho _{t} {%
\sigma}_{+} {\sigma}_{-}-2 {\sigma}_{-}\rho _{t} {\sigma}_{+}\right) ,
\end{equation}%
where $\epsilon _{t}$=Im$(\dot{b}_{t}/b_{t})$ and $\gamma _{t}$=Re$(\dot{b}%
_{t}/b_{t})$ are time-dependent Lamb shift and decay rate respectively, and $%
b_{t}$ is the decoherence function depending on certain reservoir structures
\cite{Book-Open}. The reduced density matrix of the atom with an initial
state $\rho =(\rho _{mn})$ (in matrix form) can be evaluated as \cite%
{Book-Open}
\begin{equation}
\rho _{t}=\Lambda _{t}\rho =\left(
\begin{array}{ll}
\rho _{11}|b_{t}|^{2} & \rho _{10}b_{t} \\
\rho _{01}b_{t}^{\ast } & 1-\rho _{11}|b_{t}|^{2}%
\end{array}%
\right) ,  \label{density M}
\end{equation}%
with $\Lambda _{t}$ the quantum map.

In this work, we consider an ideal photonic crystal with isotropic photon
dispersion relation approximated by $\omega _{k}$ $=\omega
_{c}+A(k-k_{0})^{2}$ near the band edge \cite{PBG}, where $\omega _{c}$ is
the upper band-edge frequency, and $A=\omega _{c}/k_{0}^{2}$ with $%
k_{0}\simeq \omega _{c}/c$ being a specific wave vector with respect to the
point-group symmetry of the PCC. The Laplace transform of $b_{t}$ is $\tilde{%
b}_{s}=[s-(i\beta )^{3/2}/\sqrt{s-i\delta }]^{-1}$ with $\beta ^{3/2}=\omega
_{0}^{7/2}d^{2}/(6\pi \epsilon _{0}\hbar c^{3})$ and $\delta =\omega
_{0}-\omega _{c}$ \cite{PBG}. The decoherence function $b_{t}$ can then be
calculated by the standard inverse Laplace transform as
\begin{equation}
\begin{array}{l}
b_{t}=\mathscr{L}^{-1}(\tilde{b}_{s}) \\
=\sum\limits_{\substack{ j<k  \\ j\neq k\neq l}}e^{-x_{l}t}\frac{%
x_{l}^{2}+i\delta x_{l}+(i\beta )^{3/2}\sqrt{-(x_{l}+i\delta )}\text{Erf}%
\left( \sqrt{-t(x_{l}+i\delta )}\right) }{(x_{j}-x_{l})(x_{k}-x_{l})}%
\end{array}%
,
\end{equation}%
where $x_{j,k,l}$ $(j,k,l=1,2,3)$ are the three parameters of the equation $%
s^{3}-i\delta s^{2}+i\beta ^{3}=(s+x_{1})(s+x_{2})(s+x_{3}),$ and Erf($\cdot
$) is the error function.

\section{Quantum speed limit time}

In this section, we evaluate the intrinsic speed for the evolution between
the initial state $\rho $ and the final state $\rho _{\tau }$, with $\tau $
the actual driving time. For convenience and without loss of generality, the
initial state is set to be the excited state $\rho =\left\vert
1\right\rangle \left\langle 1\right\vert $ \cite{op3}. It is readily checked
that the maximum in Eq. (\ref{QSL time}) is $\tau _{\infty },$ for $E_{\tau
}^{\infty }=E_{\tau }^{1}/2=E_{\tau }^{2}/\sqrt{2}$. In the light of Eqs. (%
\ref{QSL time}) and (\ref{density M}), the QSL time of the above model can
be derived as
\begin{equation}
\tau _{\mathrm{QSL}}=\frac{1-P_{\tau }}{\frac{1}{\tau }\int_{0}^{\tau
}|\partial _{t}P_{t}|dt},  \label{QSL1}
\end{equation}%
with $P_{t}=|b_{t}|^{2}$ denoting the population of excited states at time $t
$.
\begin{figure}[tbp]
\centering
\includegraphics[width=3.4in]{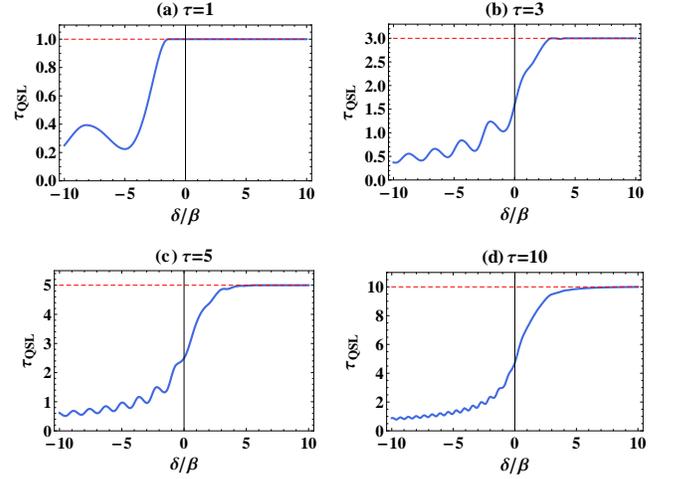}
\caption{(Color online) Quantum speed limit time $\protect\tau _{\mathrm{QSL}%
}$ in unit of $1/\protect\beta $ (blue solid curves) for an atom embedded in
an ideal isotropic photonic crystal cavity as a function of $\protect\delta /%
\protect\beta ,$ the detuning between the atom transition frequency and the
upper band-edge frequency, for different driving time $\protect\tau =1,3,5,$
and $10$ (in unit of $1/\protect\beta $) (red dashed lines) in $(a)\sim (d)$%
, respectively.}
\end{figure}

In Fig. 1, we depict the QSL time $\tau _{\mathrm{QSL}}$ in unit of $1/\beta
$ (blue solid curves) together with different actual driving time $\tau $
(red dashed lines) as functions of $\delta /\beta ,$ where the transition
from no-speedup to speedup is clearly shown. When the atom transition
frequency goes far outside the band gap (e.g., $\delta /\beta =10$), the QSL
time is actually the driving time. However, a remarkable phenomenon appears
when the atom transition frequency is approaching and going inside the band
edge ($\delta /\beta \rightarrow 0$), the QSL time $\tau _{\mathrm{QSL}}$
will begin to decrease, implying the intrinsic speedup of quantum evolution.
Obviously, the deeper the atomic transition frequency lies in the band edge,
the smaller the QSL time will be in a general trend. On the other hand, if
the driving time is not very long, e.g., in Fig. 1(a), the speedup
phenomenon only occurs inside the band edge ($\delta /\beta <0$), which is
quite different from Fig. 1(b)$\sim $(d), where the speedup region lies even
outside the band edge.

How can we explain the above phenomena? Can we claim that the reason for the
speedup in the above model is solely due to the memory (non-Markovian)
effect of the reservoir? To answer these questions, in the following we
first describe the non-Markovian behavior of the evolution of this model.

\section{Mechanism for the speedup of quantum evolution}

\begin{figure}[tbp]
\centering
\includegraphics[width=3.4in]{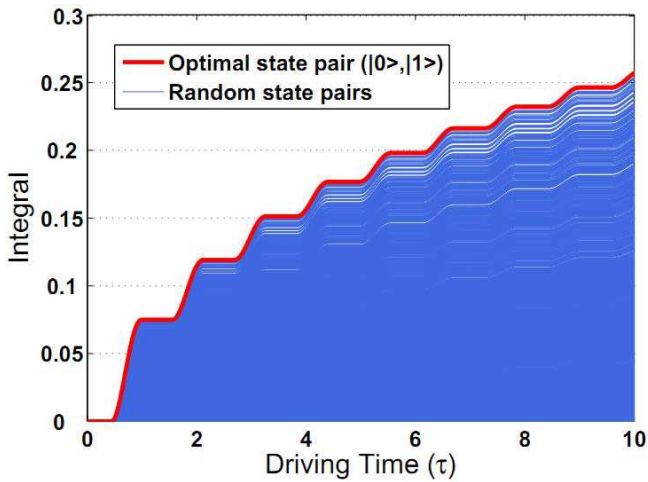}
\caption{(Color online) The integrals $\protect\int_{G_{t}>0}G_{t}dt$ in the
definition of non-Markovianity [Eq. (\protect\ref{BLP})] versus the driving
time $\protect\tau $ in the case of $\protect\delta /\protect\beta =-10$. The
red (dark gray) and green (light gray) curves represent the integrals with
the optimal state pair $(|0\rangle ,|1\rangle )$ and other 2000 randomly
generated pairs, respectively.}
\end{figure}

The non-Markovianity measure we employ here is based on the total amount of
information, quantified by trace distance $D(\Lambda _{t}\rho _{1},\Lambda
_{t}\rho _{2})=\left\Vert \Lambda _{t}\rho _{1}-\Lambda _{t}\rho
_{2}\right\Vert _{1}/2$ of a pair of evolved quantum states $(\rho _{1},\rho
_{2})$, flowing back from the environment. The gradient of trace distance $%
G_{t}=\partial _{t}D(\Lambda _{t}\rho _{1},\Lambda _{t}\rho _{2})$
represents the information flow, with positive value indicating information
flowing back to the system. Here $\Lambda =\{\Lambda _{t}\}_{t\in \lbrack
0,\tau ]}$ denotes the dynamical map. The non-Markovianity is defined as the
total backflow of information \cite{BLP}
\begin{equation}
\mathcal{N}(\Lambda )=\underset{\rho _{1},\rho _{2}}{\max }%
\int_{G_{t}>0}G_{t}dt,  \label{BLP}
\end{equation}%
with the maximization over all initial state pairs $(\rho _{1},\rho _{2})$.
There exists no general analytical method to find the optimal initial state
pair $(\rho _{1},\rho _{2})$ \cite{optimal}. Therefore, we will use
numerical calculations instead.

\begin{figure}[tbp]
\centering
\includegraphics[width=3.4in]{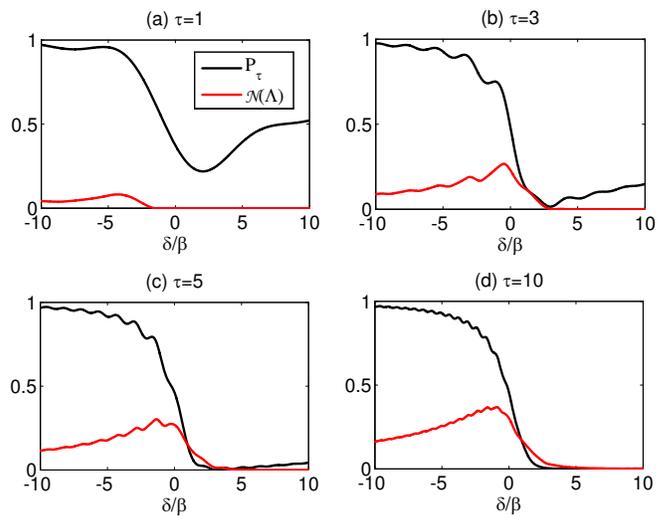}
\caption{(Color online) Non-Markovianity $\mathcal{N}(\Lambda )$ [red (dark
gray) curves] and excited population $P_{\protect\tau }=|b_{\protect\tau %
}|^{2}$ [light blue (light gray) curves] and as a function of $\protect%
\delta /\protect\beta $ in the case of $\protect\tau =1,3,5,10$ (in unit of $%
1/\protect\beta $) in $(a)\sim (d)$ respectively$.$}
\end{figure}

In Fig. 2, the integrals $\int_{G_{t}>0}G_{t}dt$ in the definition of $%
\mathcal{N}(\Lambda )$ of 2000 random initial
state pairs $(\rho _{1},\rho _{2})$ [green (light gray) curves] are generated with $\delta /\beta =-10$
as an example. Clearly all these pairs yield smaller values than that of the
state pair $(\left\vert 0\right\rangle ,\left\vert 1\right\rangle )$ [red (dark gray) curve]. It is easy to check that for this optimal state
pairs, the trace distance of the evolved states can be written as $%
D_{t}=\left\vert b_{t}\right\vert ^{2}$ $=P_{t}$ ($D_{t}$ denotes the
optimal trace distance in the following). Based on this, we take the
non-Markovianity as
\begin{equation}
\mathcal{N}(\Lambda )=\int_{\partial _{t}P_{t}>0}\partial _{t}P_{t}dt.
\label{nonM}
\end{equation}%
To associate the non-Markovianity with the QSL time, we rewrite Eq. (\ref%
{nonM}) as
\begin{equation}
\mathcal{N}(\Lambda )=\frac{1}{2}\left[ \int_{0}^{\tau }\left\vert \partial
_{t}P_{t}\right\vert dt+P_{\tau }-1\right] ,  \label{nonM2}
\end{equation}%
Consequently, the QSL time is reduced to
\begin{equation}
\tau _{\mathrm{QSL}}=\frac{\tau }{2\frac{\mathcal{N}(\Lambda )}{1-P_{\tau }}%
+1}.  \label{QSL-monM}
\end{equation}%
Clearly, the QSL time is related to two quantities: the non-Markovianity $%
\mathcal{N}(\Lambda )$ within the driving time and the atomic excited
population $P_{\tau }$. As an illustration, the two quantities $\mathcal{N}%
(\Lambda )$ [red (dark gray) curves] and $P_{\tau }$ [light blue (light
gray) curves], as functions of $\delta /\beta ,$ are drawn in Fig. 3.

Equation (\ref{QSL-monM}) implies that the transition point from no-speedup
to speedup of quantum evolution is just the point when the Markovian
environment becomes non-Markovian, for $\tau _{\mathrm{QSL}}=\tau $ when $%
\mathcal{N}(\Lambda )=0.$ It is therefore easy to account for the phenomenon
that the speedup only takes place within the bandgap edge when $\tau =1$
[Fig. 1(a)], for the transition point from Markovian to non-Markovian
environment just occurs inside the bandgap edge when the driving time is
short [Fig. 3(a)].

Equation (\ref{QSL-monM}) also provides us with a route to explore the
mechanism for the intrinsic speedup of quantum evolution. For illustration,
we consider the driving time $\tau =10$ as an example [Fig. 3(d)]. When the
atom transition frequency is far outside the bandgap edge frequency (e.g., $%
\delta /\beta =10$), the population will not be trapped, i.e., $P_{\tau
=10}\simeq 0,$ and the non-Markovianity also approaches zero, therefore, $%
\tau _{\mathrm{QSL}}=\tau $. When the atomic transition frequency is around
the bandgap edge (e.g., $\delta /\beta \in \lbrack -1,0]$), $\tau _{\mathrm{%
QSL}}$ is dependent on both the population $P_{\tau }$ and non-Markovianity $%
\mathcal{N}(\Lambda ),$ for the atomic excited population is trapped, i.e., $%
P_{\tau =10}\neq 0,$ and the non-Markovianity is strong. Finally, if the
atomic transition frequency is deeply inside the bandgap edge (e.g., $\delta
/\beta =-10$), the non-Markovianity becomes small, and $\tau _{\mathrm{QSL}}$
is mainly dependent on $P_{\tau }$. Summarizing, it is the competition
between $P_{\tau }$ and $\mathcal{N}(\Lambda )$ that takes responsibility
for the intrinsic speedup of quantum evolution.

We note that our conclusion is not only restricted to the above physical
model, it can also be used to explain the phenomenon illustrated in Ref.
\cite{op3}, where the optimal trace distance for non-Markovianity of a
two-level atom resonantly coupled to a leaky single model cavity is found to
be $D_{t}=|b_{t}|=\sqrt{P_{t}}$ \cite{xzy}. Since the monotonicity of $\sqrt{%
P_{t}}$ and $P_{t}$ is the same, for simplicity, we still use $P_{t}$ as the
trace distance (not optimal) for non-Markovianity $\mathcal{\tilde{N}}%
(\Lambda )$. Therefore Eq. (\ref{QSL-monM}) still holds with $\mathcal{N}%
(\Lambda )$ replaced by $\mathcal{\tilde{N}}(\Lambda ).$ It is easy to check
that no population will be trapped, i.e., $P_{\tau }=0$ when $\tau
\rightarrow \infty .$ Thus
\begin{equation}
\tau _{\mathrm{QSL}}=\frac{\tau }{2\mathcal{\tilde{N}}(\Lambda )+1},
\label{QSL-n}
\end{equation}%
which is inversely proportional to non-Markovianity $\mathcal{\tilde{N}}%
(\Lambda )$ if $\tau $ is longer enough. In this case, when $\mathcal{\tilde{%
N}}(\Lambda )=0$ (Markovian), $\tau _{\mathrm{QSL}}=\tau .$ Consequently,
the non-Markovian effect becomes the unique reason for speeding up quantum
evolution in this model.

\section{Possible experimental realization}

\begin{figure}[tbp]
\centering
\includegraphics[width=3.4in]{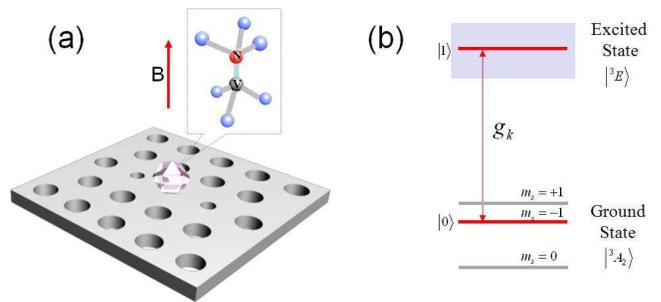}
\caption{(Color online) (a). Schematic diagram for the planar PCC-N-V
system. (b). Energy levels of the electronic spin where $\left\vert
0\right\rangle =\left\vert E_{0},m_{s}=-1\right\rangle $ and $\left\vert
1\right\rangle =(\left\vert E_{-},m_{s}=+1\right\rangle +\left\vert
E_{+},m_{s}=-1\right\rangle )/\protect\sqrt{2}$ are employed as the
two-level system coupled to the modes of the PCC with coupling constant g$%
_{k}.$}
\end{figure}

In this section, we briefly discuss a possible experimental realization for
testing the above phenomena using a composite isotropic planar PCC-N-V
system [shown in Fig. 4(a)] \cite{NV-PBG}, where an external magnetic field
\textbf{B} is applied along the [111] axis of the N-V center to lift the
degeneracy between the states $m_{s}=\pm 1.$ In our case, the ground state $%
\left\vert 0\right\rangle =\left\vert E_{0},m_{s}=-1\right\rangle $ and the
excited state $\left\vert 1\right\rangle =(\left\vert
E_{-},m_{s}=+1\right\rangle +\left\vert E_{+},m_{s}=-1\right\rangle )/\sqrt{2%
}$ are employed as the two-level open system coupled to the PCC with $\sigma
^{+}$ circular polarization mode, where $E_{0,\pm }$ refers to the orbital
state with 0,$\pm $ orbital angular momentum projection \cite%
{NV-level,NV-ywl}. In experiment, the initial state of N-V center can be
prepared by 532nm laser pulse with high fidelity \cite{NV-initial,NV-review}%
, and the finial state can be tested with standard tomography technique \cite%
{Book QIC}. To control the transition frequency in order to observe the
transition from no-speedup to speedup phenomenon, we may shift the frequency
$\omega _{0}$ with Stark effect. For the relevant experimental parameters,
the optical transition dipole moment and the typical transition frequency of
the N-V center are observed to be $d\sim 4\times 10^{-30}\mathrm{C}\cdot
\mathrm{m}$ \cite{book-diamond,NV-d} and $\omega _{0}\sim 2.9\mathrm{PHz}$
\cite{NV-property}, which lead to $\beta \sim 1\mathrm{GHz}.$ We can tune
the energy levels by the Stark shift with $\Delta \sim 5\mathrm{GHz}$ \cite%
{NV-stark}$.$ Therefore, $\Delta /\beta \simeq 5,$ which is large enough to
obverse the no-speedup to speedup transition (e.g., in Fig. 1(b), $\delta
/\beta :5\rightarrow 0$). On the other hand, numerous experiments have
successively demonstrated the strong coupling between N-V centers and the
modes of gallium phosphide PCC \cite{NVPC1}, silicon nitride PCC \cite{NVPC2}%
, and PCC in monocrystalline diamond \cite{NVPC3}, respectively. These
advances imply that this composite system is a suitable platform to test the
results.

\section{Discussion and Conclusion}

Though our present work mainly focuses on a specific open system model,
further study on more complicated non-Markovian systems characterized by
other non-Markovian master equations \cite{Book-Open}, e.g., post-Markovian
master equation \cite{post-nonM}, will be of great interest and importance.

On the other hand, as the non-Markovian effect plays a significant role in
the speedup of quantum evolution, one interesting question then arises: will
non-Markovianity also be associated with some other speed limits? In quantum
many-body systems, for instance, there exists a maximum speed of information
propagation in discrete quantum systems with local interactions. This speed
limit, known as Lieb-Robinson bound \cite{LR}, has been extensively studied
\cite{LR1} and was observed in experiment with a one-dimensional ultracold
gas of bosonic atoms \cite{LR-exp}. Since quantum systems are always subject
to decoherence and dissipation in practice, recent studies of Lieb-Robinson
bound have been extended to classical Markovian dynamics \cite{LR-Mc} as
well as quantum Markovian \cite{LR-M} and non-Markovian \cite{LR-nonM}
conditions. It is desirable to further investigate the effects of noise and
memory environments on the Lieb-Robinson velocity bound, and the interplay
between QSL and the Lieb-Robinson bound.

In summary, for a model of two-level atom embedded in a PBG structured
reservoir, the transition from no-speedup to speedup of quantum evolution
has been observed. In particular, we have established a linkage between the
QSL time and non-Markovianity as well as the population of excited states
for a given driving time. We have found that it is the competition between
the two quantities that finally determines the speedup of quantum evolution
under noise. The phenomenon we illustrated in this work is finally analyzed
by real PCC-N-V based experimental data. Our approach may be of both
theoretical and experimental interests in exploring the ultimate limits to
quantum computers in memory environments.

\section*{ACKNOWLEDGMENTS}

Z.-Y.X. thanks Dr. Sebastian Deffner for helpful comments. This work was
supported by NNSFC under Grant Nos. 11204196, 11274351, 11074184, 11375259,
61134008, SRFDPHE under Grant No. 20123201120004, and NCMIS-CAS under Grant
No. Y029152K51.

\end{document}